\documentclass[12pt]{iopart}
\usepackage{graphicx}

\begin{document}

% A useful Journal macro
\def\Journal#1#2#3#4{{#1} {\bf #2}, #3 (#4)}

% Some useful journal names
\def\NCA{Nuovo Cimento}
\def\NIM{Nucl. Instr. Meth.}
\def\NIMA{{Nucl. Instr. Meth.} A}
\def\NPB{{Nucl. Phys.} B}
\def\NPA{{Nucl. Phys.} A}
\def\PLB{{Phys. Lett.}  B}
\def\PRL{Phys. Rev. Lett.}
\def\PRC{{Phys. Rev.} C}
\def\PRD{{Phys. Rev.} D}
\def\ZPC{{Z. Phys.} C}
\def\JPG{{J. Phys.} G}
\def\CPC{Comput. Phys. Commun.}
\def\EPJ{{Eur. Phys. J.} C}

\title[Physics with Identified Particles at STAR]{Physics with Identified Particles at STAR}

\author{Lijuan Ruan (for the STAR Collaboration\footnote[1]{For the full author list
and acknowledgements see Appendix ``Collaborations'' in this
volume.})}

\address{Nuclear Science Division, MS 70R319, Lawrence Berkeley National Lab, Berkeley, CA 94720}
\ead{ljruan@lbl.gov}
\begin{abstract}
New physics results with identified particles at STAR are
presented. Measurements at low $p_T$ address bulk properties of
the collision, while those at high $p_T$ address jet energy loss
in the bulk matter produced. Between these extremes, measurements
at intermediate $p_T$ address the interplay between jets and the
bulk. We highlight: measurements of $v_2$ fluctuations as a new,
sensitive probe of the initial conditions and the equation of
state; correlations involving multi-strange particles, along with
ratios of identified particles to test coalescence as a mechanism
of particle production at intermediate $p_T$; three particle
azimuthal correlation to search for conical emission; and the
energy and particle-type dependence of hadron production at high
$p_T$ to study quark and gluon jet energy loss.
\end{abstract}

%\maketitle

\section{Introduction}
Comparisons between experiment and theory suggest that central
Au+Au collisions at RHIC produce dense and rapidly thermalizing
matter characterized by initial energy densities far above the
critical values predicted by lattice QCD for formation of a
Quark-Gluon Plasma, accompanied by nearly ideal fluid flow, marked
by constituent interactions of very short mean free path,
established most probably at a stage preceding hadron formation,
along with opacity to jets~\cite{starwhitepaper}. In order to
study the properties of the partonic matter created in Au+Au
collisions in detail, we would like to address the following
physics topics using the data taken by the STAR experiment at
RHIC:

Bulk properties and Equation of State (EOS) by using anisotropic
flow at low $p_T$ ($p_T <$ 2 GeV/c), energy dependences of
fluctuations and freeze-out properties, and intermediate and low
$p_T$ charged hadron azimuthal correlations.

Coalescence/recombination at intermediate $p_T$ (2 $< p_T <$ 6
GeV/c) by using identified baryon and meson $v_2$, baryon and
charged hadron azimuthal correlations, and energy dependence of
particle production.

Parton energy loss in the medium by using identified particle
spectra at high $p_T$ ($p_T >$ 6 GeV/c).

\section{Equation of State}

\subsection{Anisotropic Flow $v_1$, $v_2$ and $v_2$ Fluctuation}
\begin{figure}[h] \centerline{
\includegraphics[width=0.34\textwidth,height=0.31\textwidth]
{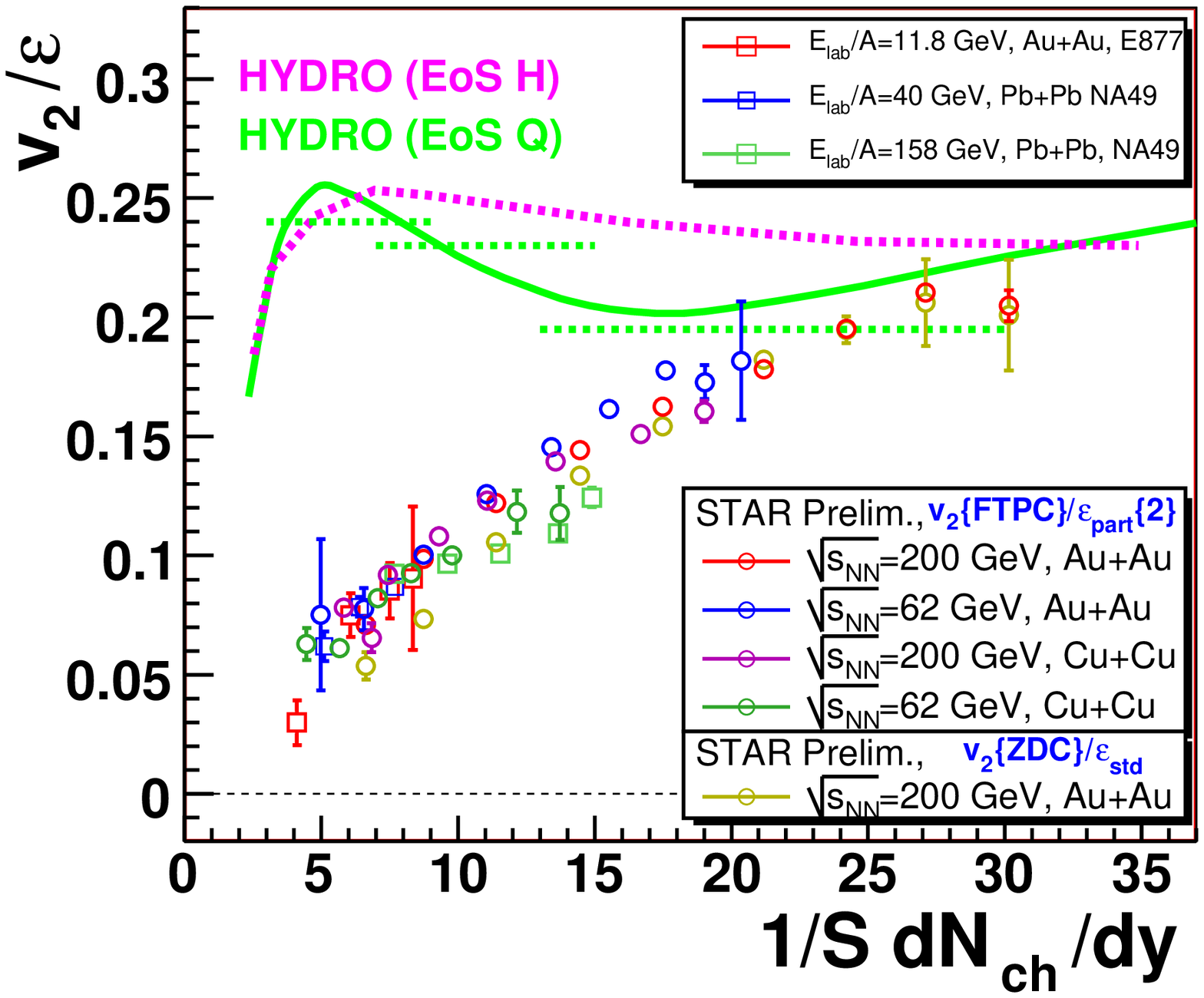}\includegraphics
[width=0.35\textwidth]{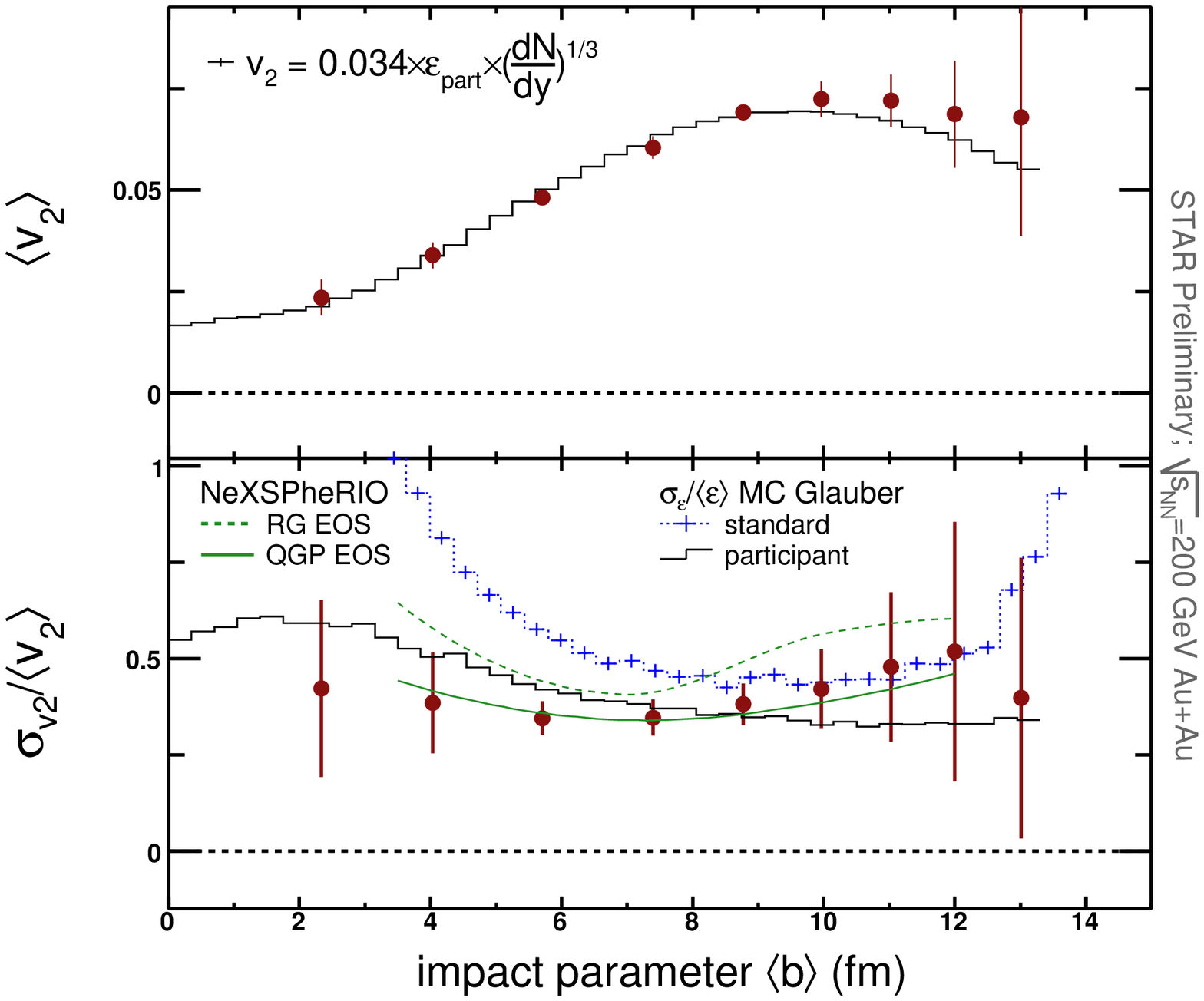}
\includegraphics
[width=0.33\textwidth,height=0.25\textwidth]{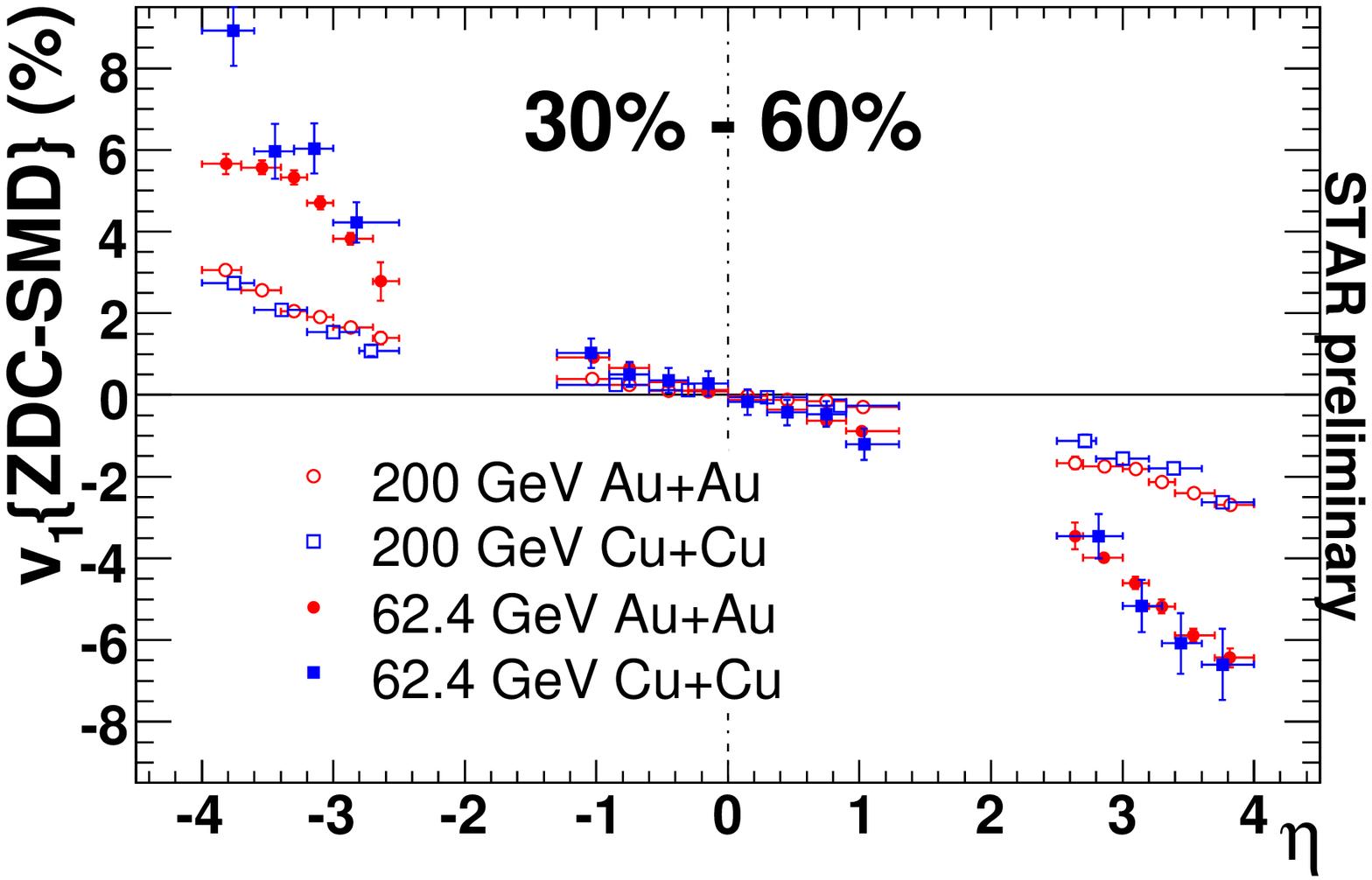}}
\vspace{-0.35cm} \caption[]{(in color on line) (left) The
$v_2/\epsilon$ versus $\frac{1}{S} dN_{ch}/dy$. (middle) The
$\langle v_2 \rangle$ and $\sigma_{v_2}/\langle v_2 \rangle$
versus $\langle b \rangle$ in Au+Au collisions at 200 GeV. (right)
$v_1$ versus pseudo-rapidity $\eta$.} \label{v2plot}
\end{figure}

The large elliptic flow ($v_2$) observed at RHIC is sensitive to
pressure gradients developed in the early stages of heavy ion
collisions, and so is believed to be sensitive to the EOS. Further
insight into the processes governing the evolution of the system,
and the physics underlying the formation of the flow itself, can
be provided by the study of its dependence on system size and
collision energy. Fig.~\ref{v2plot} (left) shows the ratio of
integral $v_2$ over the spatial eccentricity ($\epsilon$) as a
function of charged particle density in the transverse plane
($\frac{1}{S} dN_{ch}/dy$) from Au+Au and Cu+Cu collisions at
$\sqrt{s_{_{NN}}}=$ 62 and 200 GeV~\cite{svoloshin}, where $S$ is
the overlap area in the transverse plane weighted with the number
of participants along the beam axis~\cite{art}. In this plot,
$v_2\{FTPC\}$ is derived via correlations between a particle in
the main TPC and those in the Forward TPC~\cite{svoloshin},
reducing non-flow effects, while $v_2\{ZDC\}$ is derived using the
reaction plane from the STAR ZDC-SMD. After $v_2\{FTPC\}$ and
$v_2\{ZDC\}$ are scaled with appropriate values of
eccentricity~\cite{svoloshin}, a universal curve of $v_2/\epsilon$
versus $\frac{1}{S} dN_{ch}/dy$ is observed for all the energy and
collision systems. The $v_2/\epsilon$ value in central Au+Au
collisions reaches the hydrodynamic limit with QGP EOS, suggesting
possible thermalization achieved in central Au+Au collisions.

The first $v_2$ fluctuation measurement at STAR allows us to
remove the major source of systematic uncertainty for $\langle v_2
\rangle$ and provides sensitivity to initial
conditions~\cite{psorensen}. Fig.~\ref{v2plot} (middle) shows, as
a function of impact parameter ($\langle b \rangle$), both the
mean $v_2$ ($\langle v_2 \rangle$) (upper panel) and a measure of
the event-by-event fluctuations of $v_2$, the r.m.s. over the mean
($\sigma_{v_2}/\langle v_2 \rangle$) (lower panel). The
fluctuation measure is found to be $\sim$ 36\%, approximately
independent of centrality, which corresponds well with the
fluctuations in initial-state eccentricity derived in a Glauber
model~\cite{psorensen}. Comparisons with hydrodynamic models
indicate further that the fluctuations may be sensitive to the
EOS. The dependence of directed flow ($v_1$) on system size is
strikingly different from that of $v_2$. While $v_2$ appears to
scale with the transverse particle density, $v_1$ appears to scale
with the fraction of the hadronic cross-section measured.
Fig.~\ref{v2plot} (right) shows that, while $v_1$ has a strong
dependence on collision energy, for the same fraction of the
hadronic cross-section, e.g. collisions with 30-60\% centrality,
$v_1$ is similar in Cu+Cu and Au+Au collisions~\cite{gwang}.

\subsection{Two and Three Particle Correlations --- Medium Responses to Jet Energy Loss}
\begin{figure}[h] \centerline{
\includegraphics[width=0.33\textwidth,height=0.25\textwidth]
{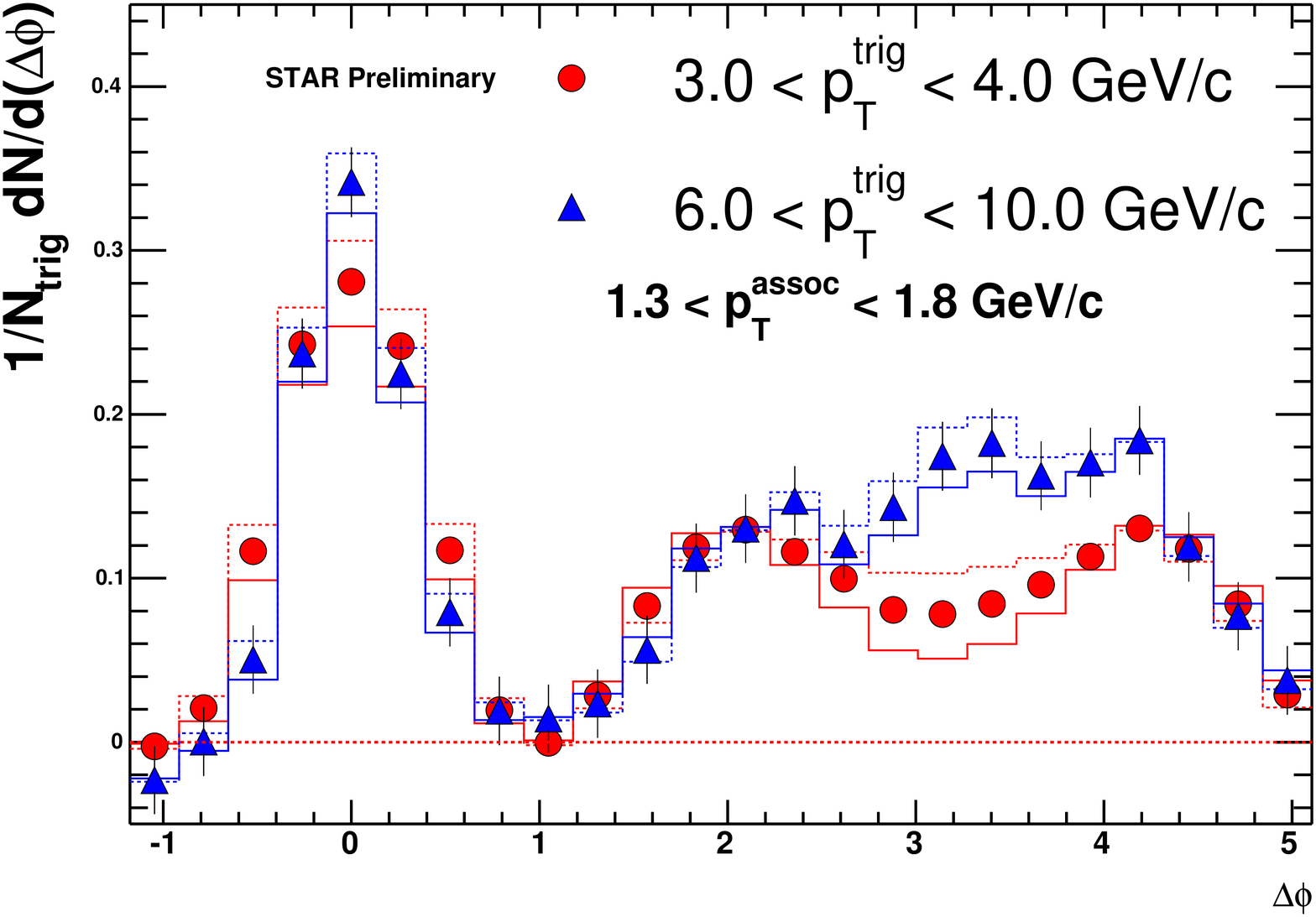}\includegraphics
[width=0.33\textwidth]{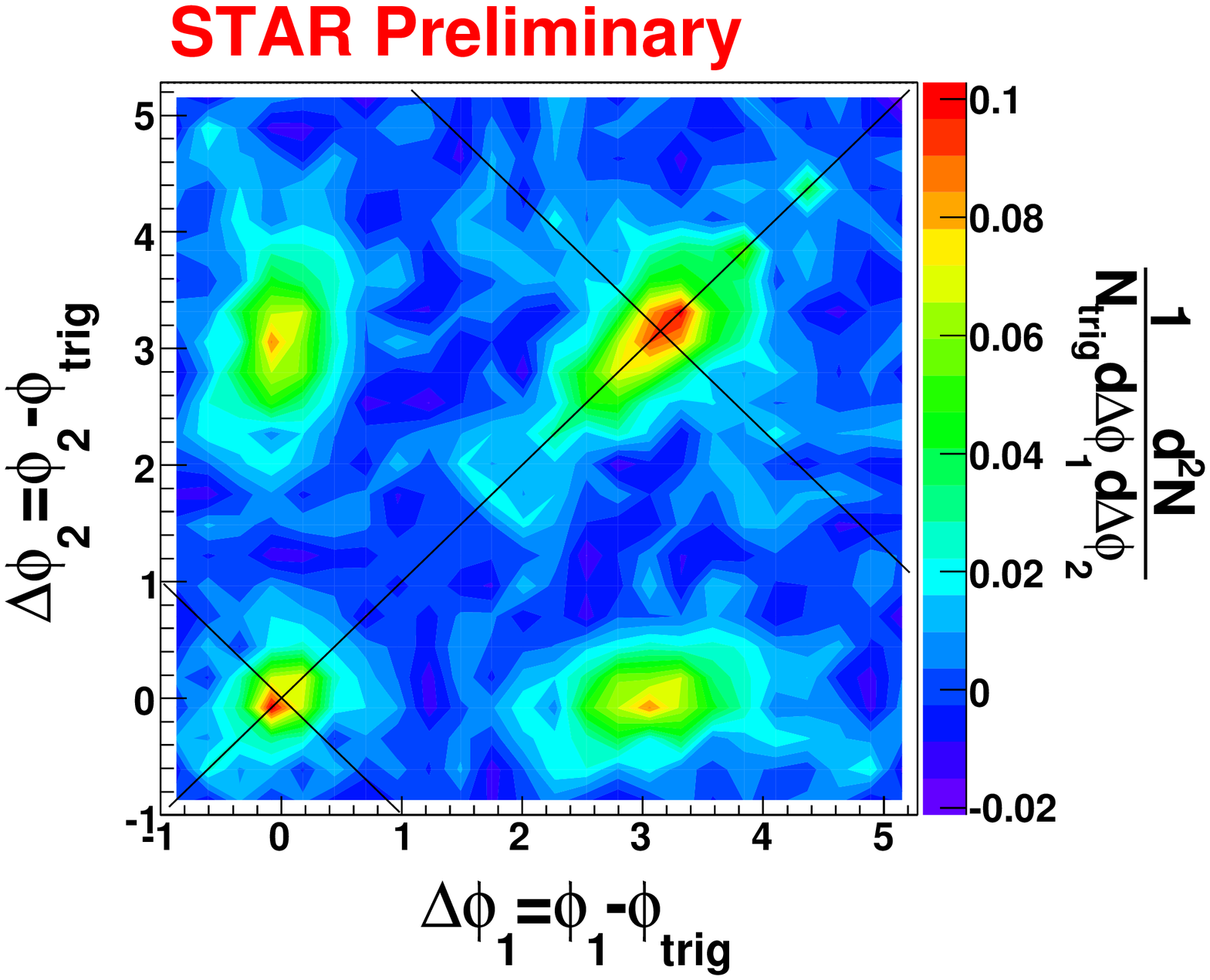}
\includegraphics
[width=0.33\textwidth]{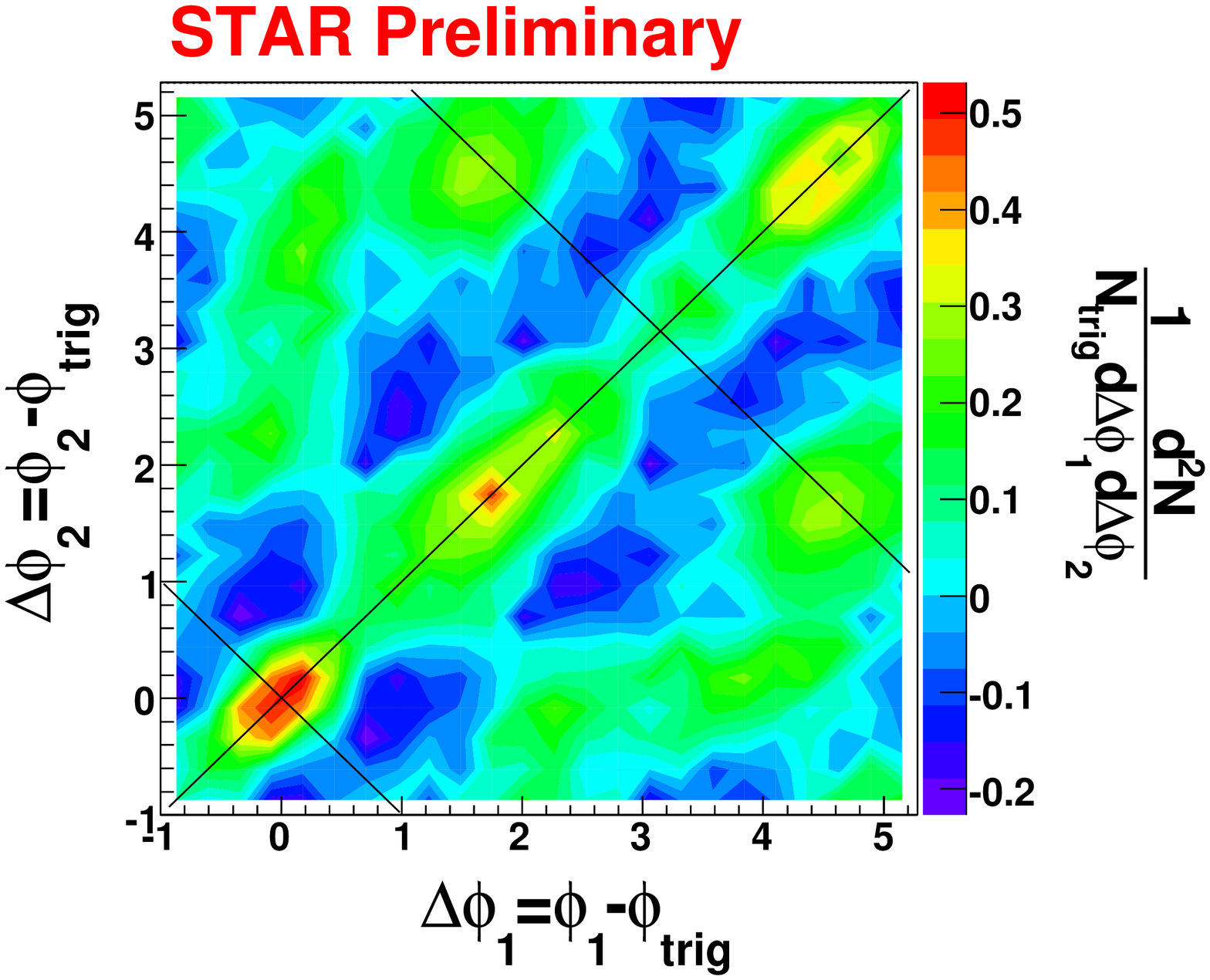}}
\vspace{-0.35cm} \caption[]{(in color on line) (left) Two particle
azimuthal correlations in 0-12\% central Au+Au collisions after
background subtraction. (middle and right) Three particle
azimuthal correlations after background subtraction in 200 GeV
d+Au and 0-12\% central Au+Au collisions respectively with a
trigger particle at 3 $< p_T^{trig} <$ 4 GeV/c and two associated
particles at 1 $< p_T^{asso} <$ 2 GeV/c.}
\label{particlecorrelation}
\end{figure}

The comparison of the measurement of elliptic flow and its
calculated value from hydrodynamic models indicates early
thermalization in central Au+Au collisions. If this is true, the
next question is how the system achieves thermalization. In
correlations between particles at high $p_T$ (hard) and those at
low $p_T$ (soft), the mean $p_T$ of soft particles on the
away-side from the hard trigger particle approaches that of the
bulk soft particles in central Au+Au collisions~\cite{starjet},
indicating that hard-soft interactions lead to partial
thermalization. One proposed mechanism for rapid thermalization
between the deposited energy and the bulk medium is through
Mach-cone shock waves, which disperse energy into collective
modes~\cite{deflectedjet,machcone}. Fig.~\ref{particlecorrelation}
(left) shows the di-hadron azimuthal correlations in central Au+Au
collisions for a fixed associated $p_T$ range and two selections
of trigger $p_T$ ($p_T^{trig}$) ranges~\cite{mhorner}. The
away-side structure strongly depends on $p_T^{trig}$. And a double
peaked structure on the away-side is observed for the lower
$p_T^{trig}$ range. This structure can be generated by several
physics mechanisms including Mach-cone shock
waves~\cite{deflectedjet,machcone}, deflected jets, and
$\check{\rm C}$erenkov radiation~\cite{cherenkov}.

In order to distinguish and/or further separate the contributions
from different physics scenarios, event-by-event three particle
correlations are measured at STAR using two analysis approaches:
the cumulant method, and background-subtraction method. The
results from the cumulant method show unambiguous evidence for
three particle correlations~\cite{cpruneau}. In order to interpret
the signal, the background-subtraction method, using a model
treating the event as the sum of correlated particles and an
uncorrelated background (with flow anisotropy), is
applied~\cite{cpruneau,julery}. Fig.~\ref{particlecorrelation}
(middle) and (right) show three particle correlations from the
background subtraction method in d+Au and central Au+Au collisions
after background subtraction~\cite{cpruneau,julery}. In these
plots, the deflected jet effect will lead to the on-diagonal
structure and Mach-cone shock waves/$\check{\rm C}$erenkov
radiation will lead to additional off-diagonal structure. In d+Au
collisions, clear near-side and away-side jet structures are
observed. In central Au+Au collisions, in addition to those,
on-diagonal and off-diagonal structures are observed. The
on-diagonal structure is consistent with the deflected jet
scenario. The off-diagonal one shows evidence for conical
emission. Whether it is due to Mach-cone or gluon $\check{\rm
C}$erenkov radiation needs more detailed quantitative comparison
with theoretical models and measurements with identified
particles, different beam energy, as well as associated $p_T$
dependence, preliminary results of which are presented at this
conference by STAR~\cite{julery}.

\subsection{Bulk Properties in Energy Scan}
\begin{figure}[h] \centerline{
\includegraphics[width=0.33\textwidth,height=0.37\textwidth]
{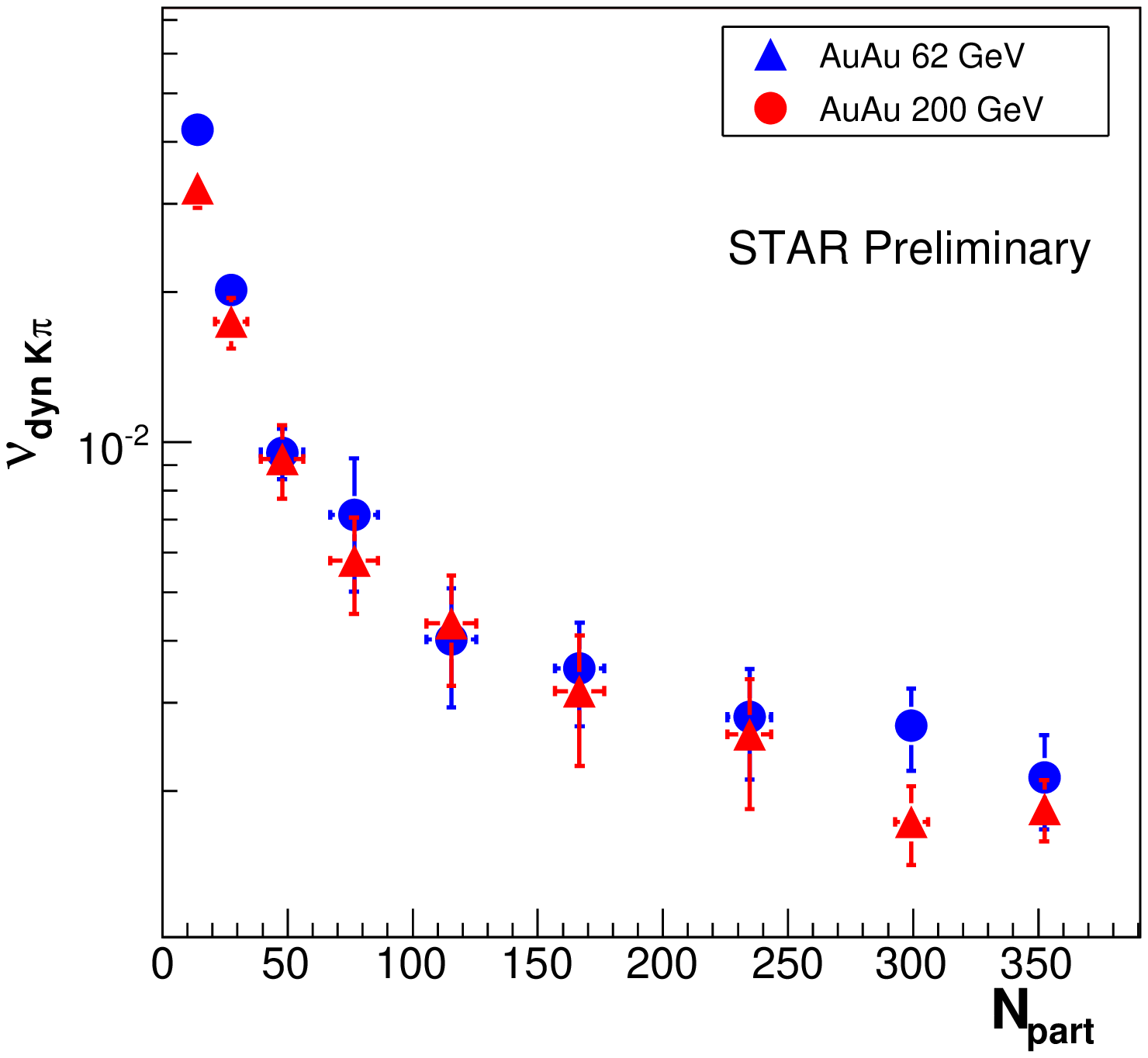}\includegraphics
[width=0.33\textwidth,height=0.4\textwidth]{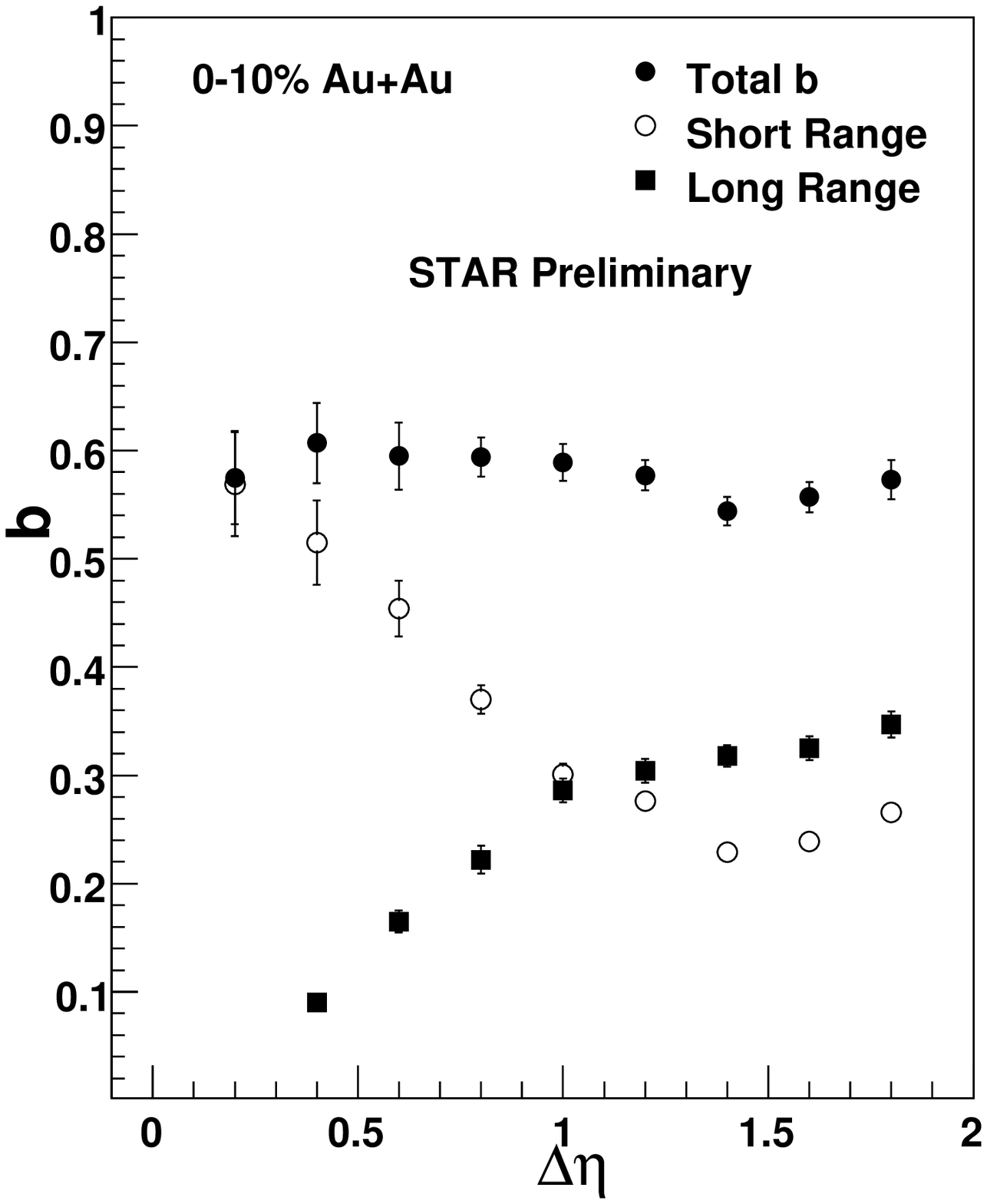}
\includegraphics
[width=0.33\textwidth,height=0.36\textwidth]{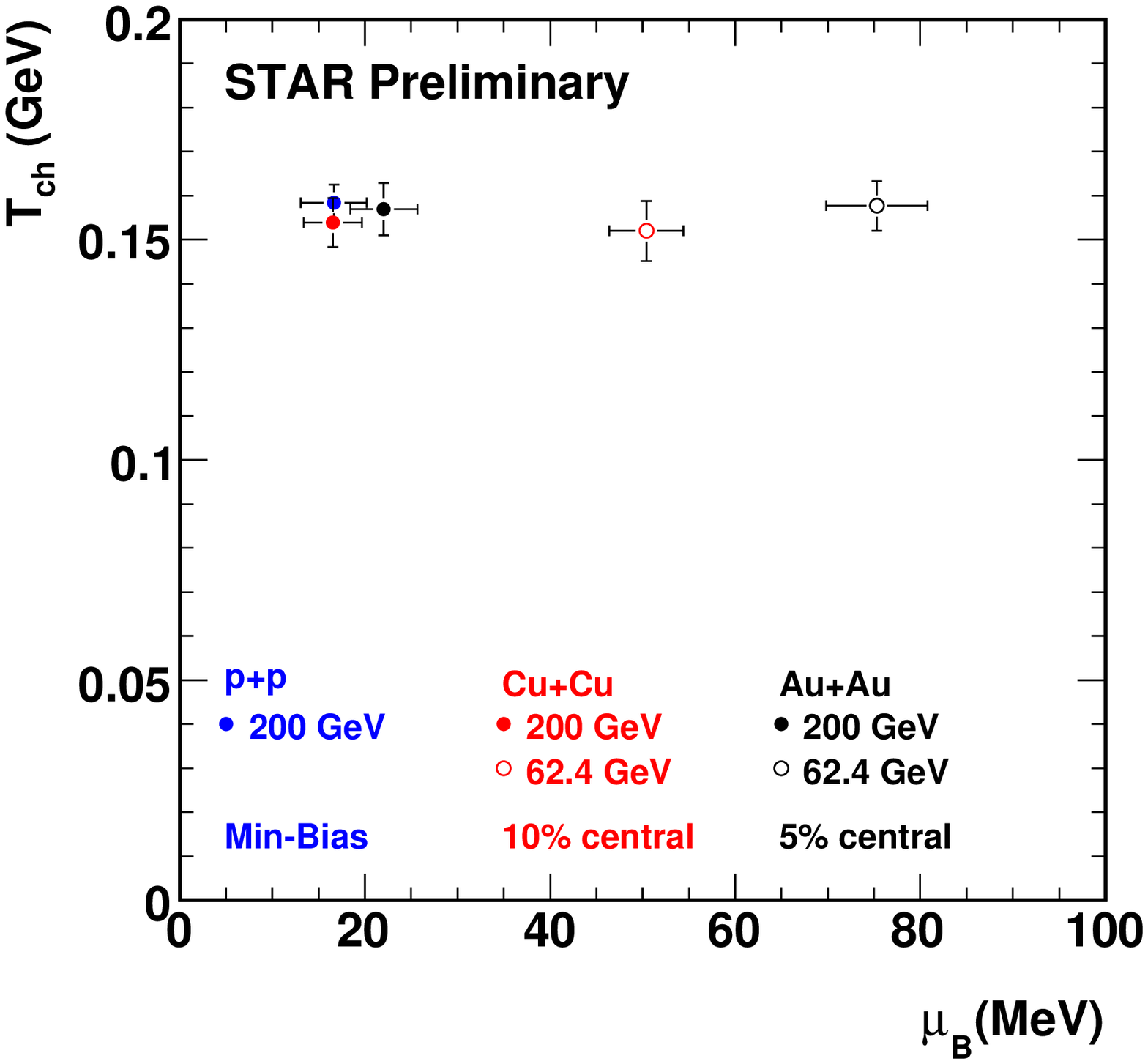}}
\vspace{-0.35cm} \caption[]{(in color on line) (left) The
$v_{dyn}$ $_{K\pi}$ fluctuation versus centrality. (middle) The
long range and short range correlation strength versus
pseudo-rapidity gap. (right) $T_{ch}$ versus $\mu_{B}$.}
\label{fluctuation}
\end{figure}

Fig.~\ref{fluctuation} (left) shows that the dynamical
fluctuations in $K/\pi$ ratio ($v_{dyn}$ $_{K\pi}$~\cite{sdas})
are similar in Au+Au collisions at 62 and 200 GeV~\cite{sdas}.
Fig.~\ref{fluctuation} (middle) shows the correlation strength
($b$) and its long and short range components, versus
pseudo-rapidity gap ($\eta_{gap}$) in central Au+Au
collisions~\cite{bsrivastava}. The growth of long range
$\Delta\eta$ correlations is clearly seen, the existence of which
is the signature of multiple parton inelastic collisions and hence
possible creation of dense partonic matter~\cite{longrange}.
Fig.~\ref{fluctuation} (right) shows the chemical freeze-out
temperature ($T_{ch}$) versus baryon chemical potential
($\mu_{B}$) derived from the thermal fit to $\pi, K$ and $p$
spectra~\cite{aiordanova}. With decreasing beam energy, the baryon
chemical potential increases. In the QCD phase diagram, the
critical point, namely the end point of the first order phase
transition, is predicted to be at 350 $< \mu_{B} <$ 700 MeV from
Lattice QCD calculations~\cite{criticalpoint}. This value is
significantly larger than the $\mu_{B}$ value at which matter is
formed in RHIC collisions in energies scanned to date. In a future
energy scan to lower energy, matter will be formed in the relevant
region in $\mu_{B}$, allowing for a search for the critical point
using systematic measurements of fluctuations and spectra of
identified particles.

%Soft physics variables such as kinetic freeze out temperature,
%mean $p_T$, HBT radii, seem to be strongly correlated with charged
%particle multiplicity~\cite{aiordanova,ddas}.

\section{Coalescence}
\subsection{Number of Constituent Quark (NCQ) Scaling of $v_2$}
\begin{figure}[h] \centerline{
\includegraphics[width=0.55\textwidth]
{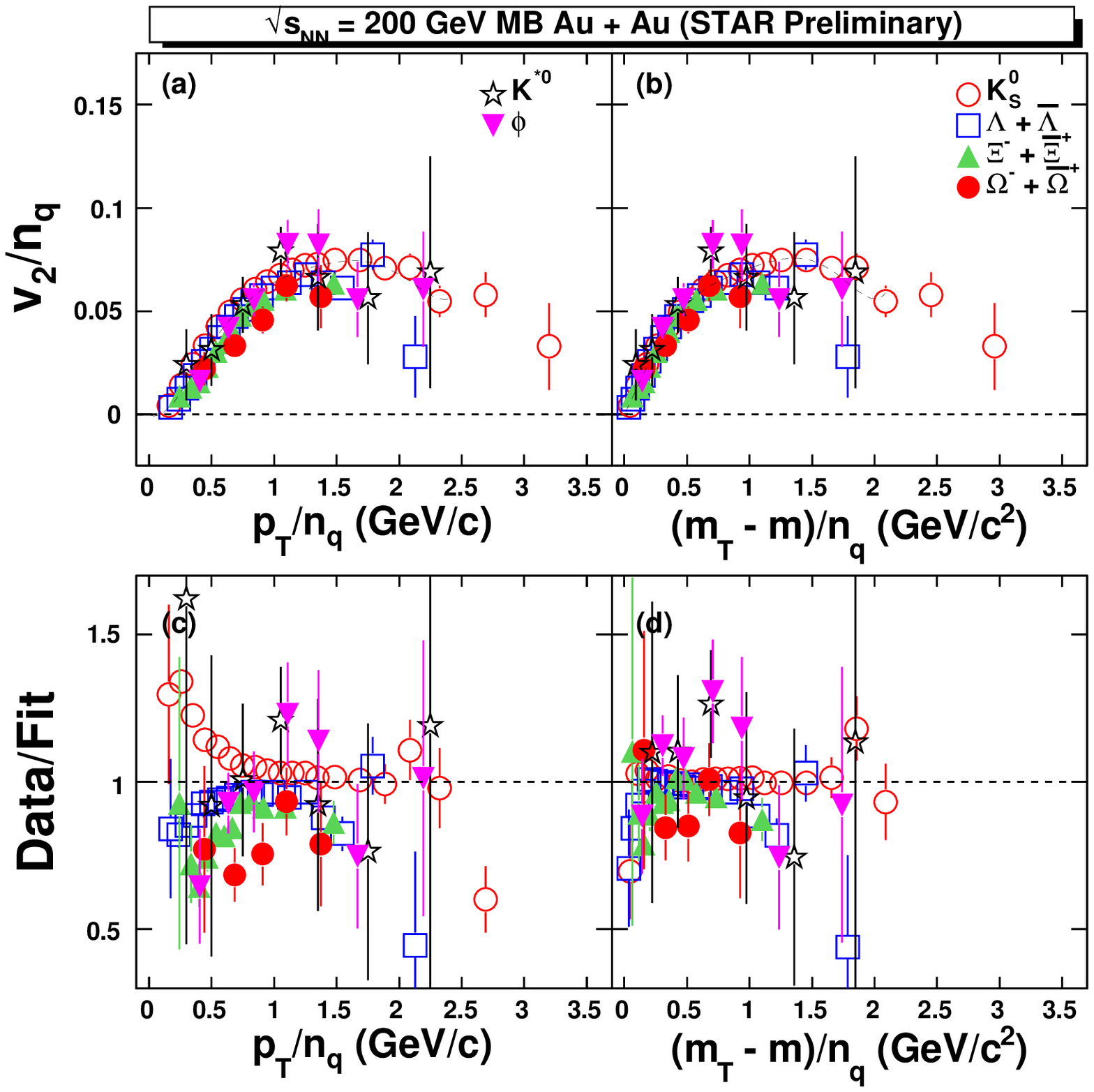}\includegraphics
[width=0.45\textwidth]{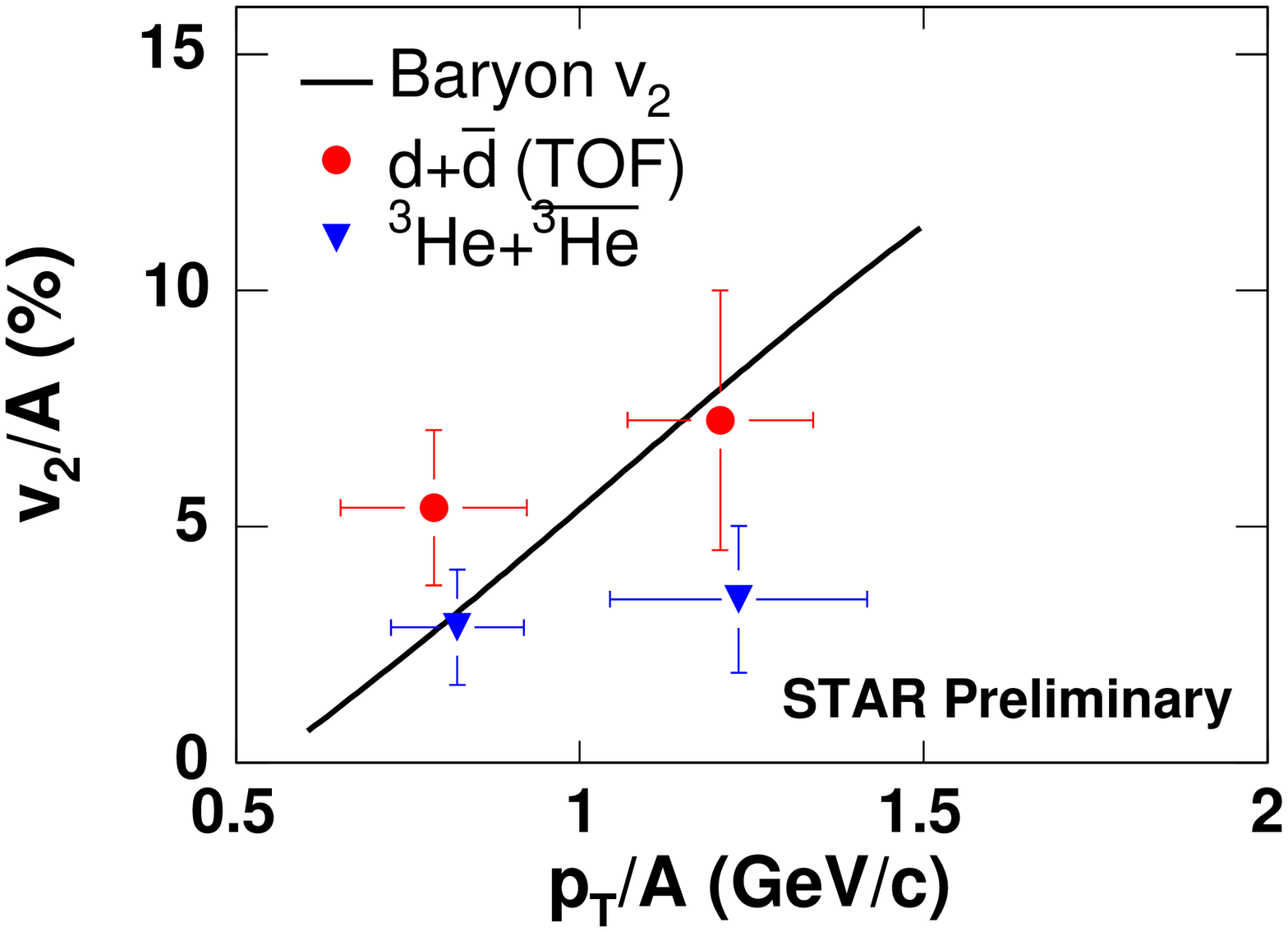} }
\vspace{-0.35cm} \caption[]{(in color on line) (left) Identified
baryon and meson $v_2/n_q$ versus $p_T/n_q$ or $(m_T-m_0)/n_q$.
(right) The atomic mass number (A) scaled $v_2$ of light nuclei
versus $p_T/A$.} \label{identifiedv2}
\end{figure}

Fig.~\ref{identifiedv2} (left) shows $v_2$ divided by the number
of constituent quarks ($n_q$) versus $p_T/n_q$ or $(m_T-m_0)/n_q$
of $K_{S}^{0}$, $\Lambda$, $\Xi$, $\Omega$, $K^*$ and
$\phi$~\cite{ybai,sblyth,xdong}. A common fit to all the data
points was used and the ratios of the data points over the common
fit function are shown in the lower panel. At low $p_T$, $v_2$
shows strong mass ordering, consistent with hydrodynamical
models~\cite{hydro}. At intermediate $p_T$, $v_2$ of mesons
($K_{S}^{0}$, $K^*$, $\phi$) and baryons ($\Lambda$, $\Xi$,
$\Omega$) follow NCQ scaling, consistent with hadronization
through coalescence of constituent quarks from a collective
partonic system~\cite{voloshin,hwa,fries,ko}. Light nuclei, such
as the deuteron and $^{3}He$, can be formed by coalescence at the
nucleon level when interactions between nucleons and other
particles are weak. The $v_2$ measurements of nuclei can be used
to test the general feature of coalescence.
Fig.~\ref{identifiedv2} (right) shows $v_2$ of deuteron and
$^{3}He$ divided by atomic mass number (A) as a function of
$p_T/A$~\cite{hliu}. The deuteron $v_2$ seems to follow A scaling
while $^{3}He$ shows a possible deviation at the higher $p_T$ bin.

\subsection{Multi-Strange Particle Production and Correlations}
\begin{figure}[h] \centerline{
\includegraphics
[width=0.45\textwidth,height=0.35\textwidth]{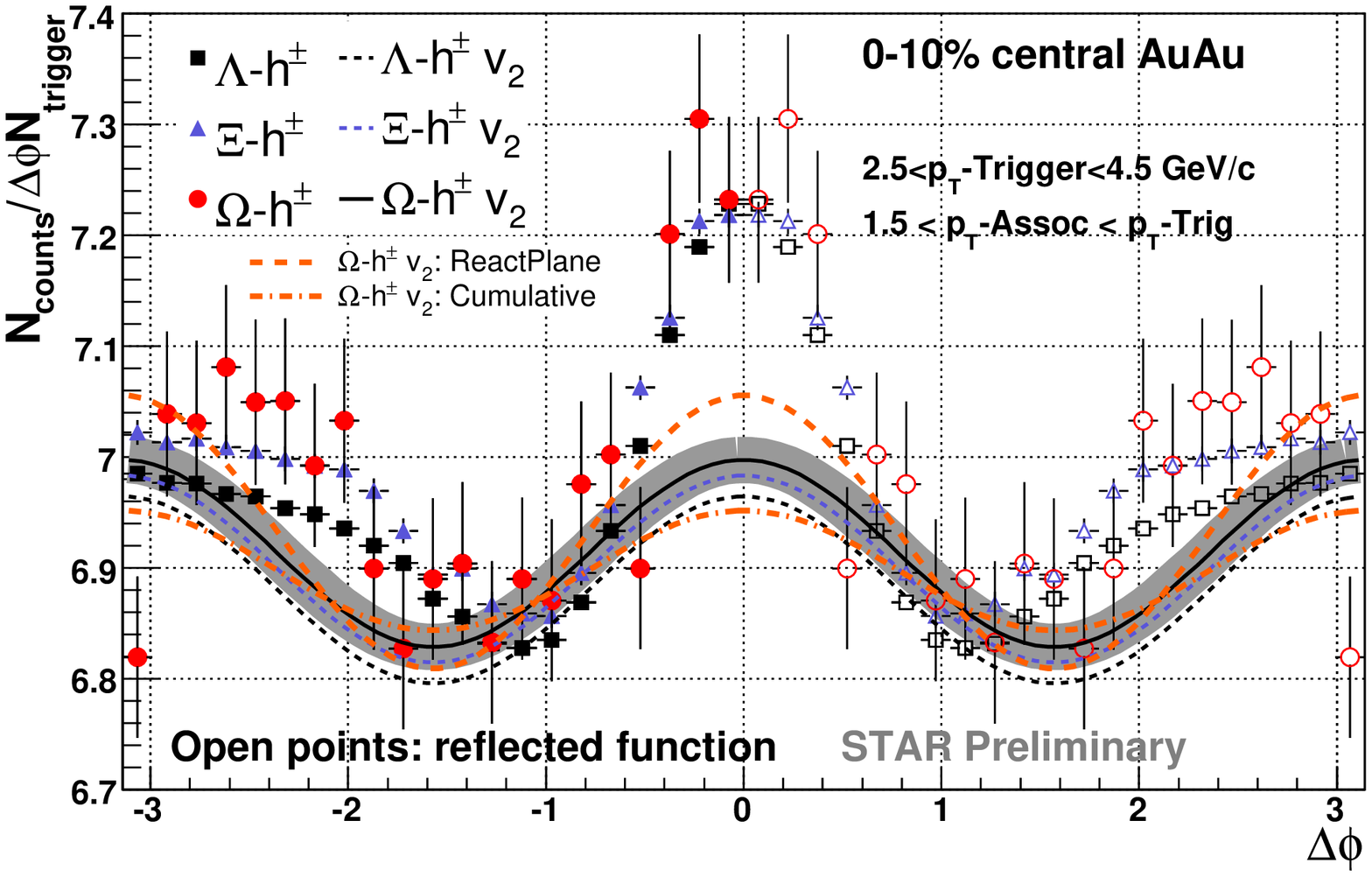}
\includegraphics
[width=0.40\textwidth]{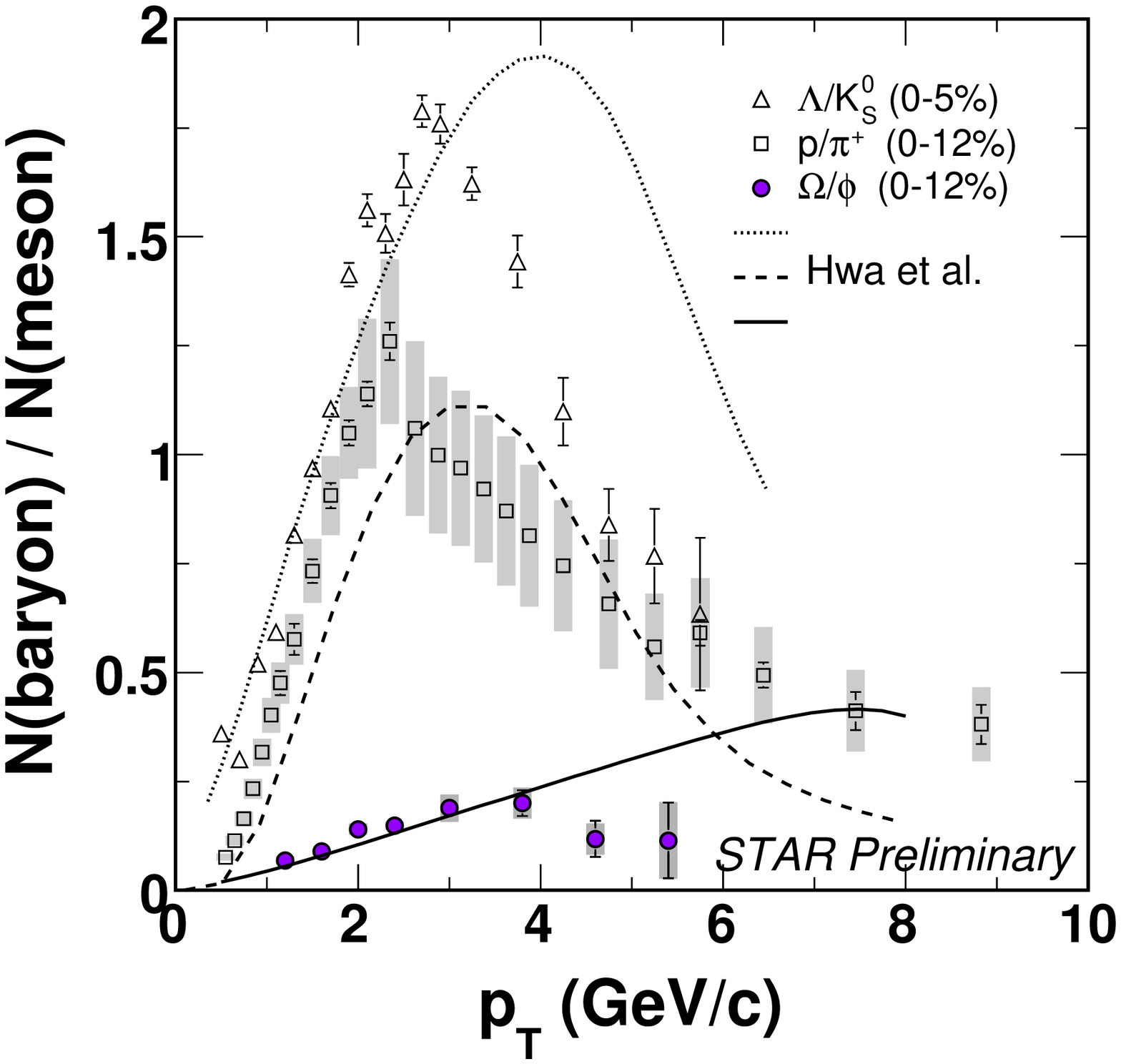}} \vspace{-0.35cm}
\caption[]{(in color on line) (left) The baryon and charged hadron
azimuthal correlations in central Au+Au collisions at 200 GeV.
(right) The baryon over meson ratios in central Au+Au collisions
at 200 GeV.}\label{omegaphi}
\end{figure}

To further understand the particle production mechanisms and test
the coalescence model predictions, we measure the correlations
between strange and multi-strange baryons and charged hadrons.
Shown in Fig.~\ref{omegaphi} (left) is the azimuthal correlation
of two particles with a baryon ($\Lambda$, $\Xi$, or $\Omega$)
trigger, and an associated charged hadron
($h$)~\cite{jbielcikova}. The $\Lambda$, $\Xi$, or $\Omega$ - $h$
correlations are similar, and a strong near-side correlation of
$\Omega-h$ is observed. This is in contrast to the recombination
model~\cite{hwastrangeness}, which predicts no near-side particle
to be associated with a multi-strange $\Omega$ or $\phi$ trigger.
In this model, the multi-strange particles ($\Omega(sss)$ and
$\phi(s\bar{s})$) mainly come from thermal s quark recombination.

Fig.~\ref{omegaphi} (right) shows the $\Omega/\phi$,
$\Lambda/K_{S}^{0}$ and $p/\pi^{+}$ ratios versus $p_T$ in 200 GeV
central Au+Au collisions~\cite{sblyth}, compared with the
calculations from a recombination model~\cite{hwa,hwastrangeness}.
The $\Omega/\phi$ ratio increases as a function of $p_T$, reaches
its maximum value at $p_T \sim$  4 GeV/c, then shows a decreasing
trend at higher $p_T$. The recombination
model~\cite{hwastrangeness} can describe the $\Omega/\phi$ ratio
up to $p_T$ $\sim$ 4 GeV/c but fails to reproduce the ratio at
higher $p_T$, indicating non-thermal s quark contribution to the
$\phi$ and $\Omega$ production in this $p_T$ region.

\subsection{Energy Dependence of Identified Particle Production at Intermediate/High $p_T$}
\begin{figure}[h] \centerline{
\includegraphics[width=0.3\textwidth,height=0.23\textwidth]
{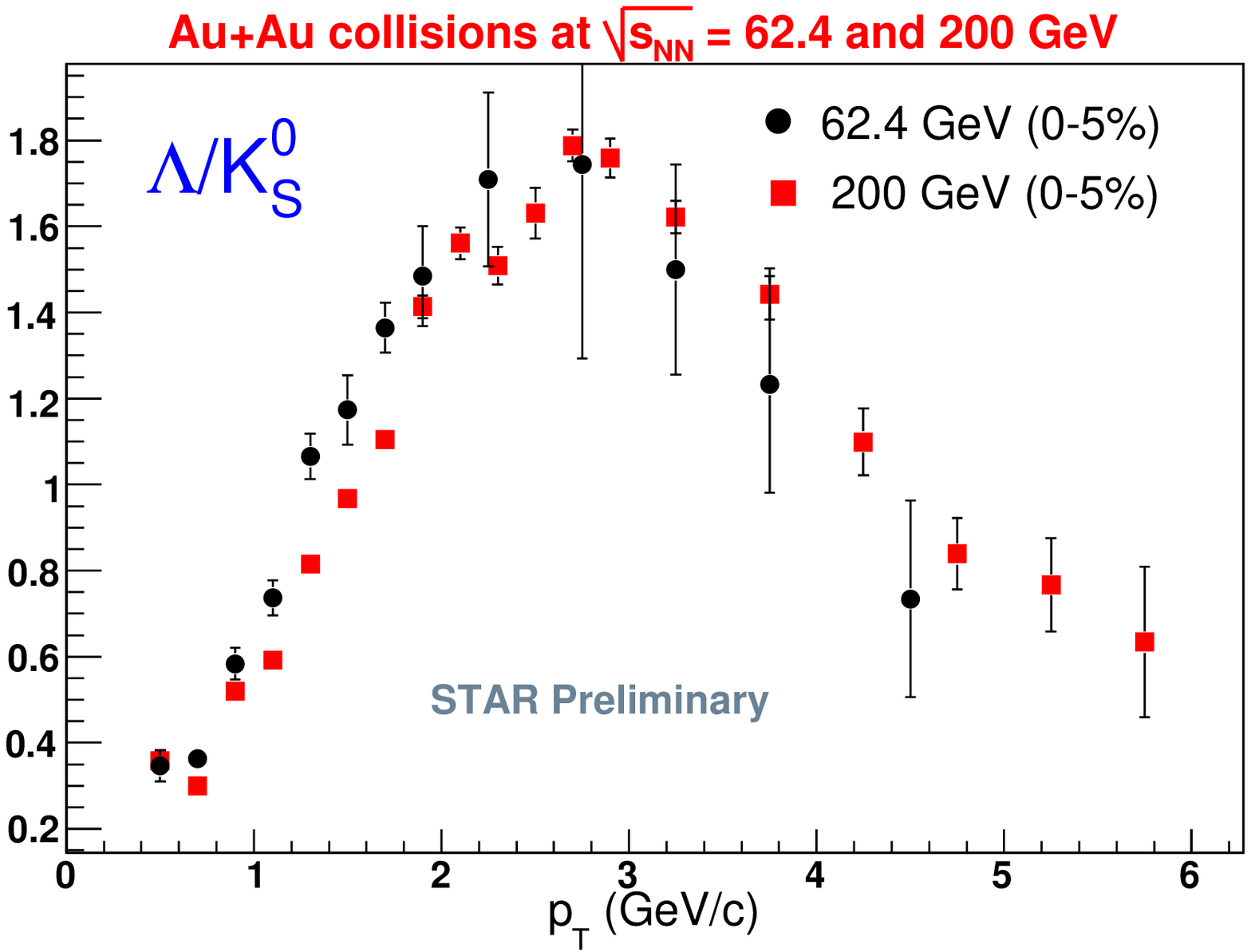}
\includegraphics[width=0.3\textwidth]
{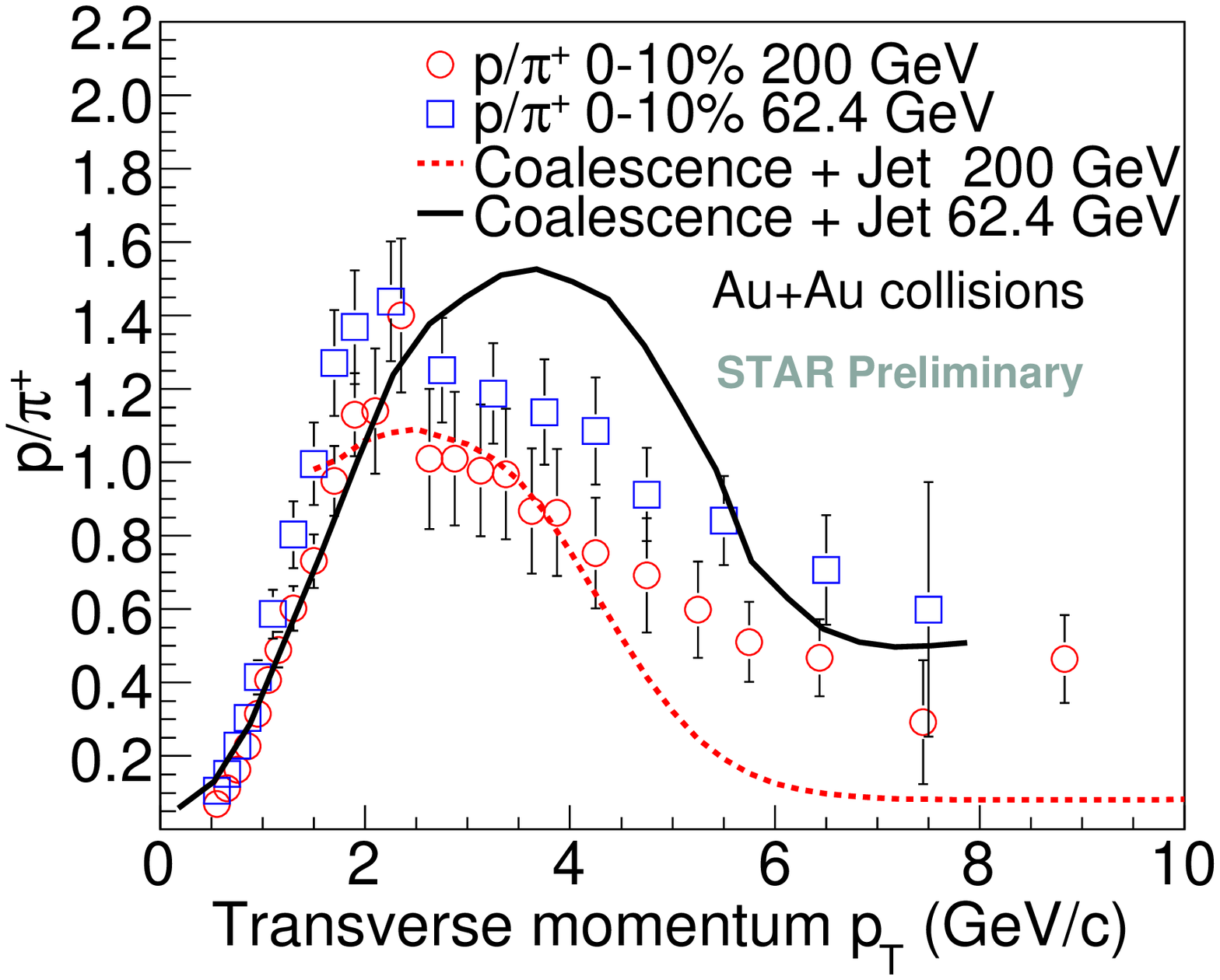}\includegraphics
[width=0.3\textwidth]{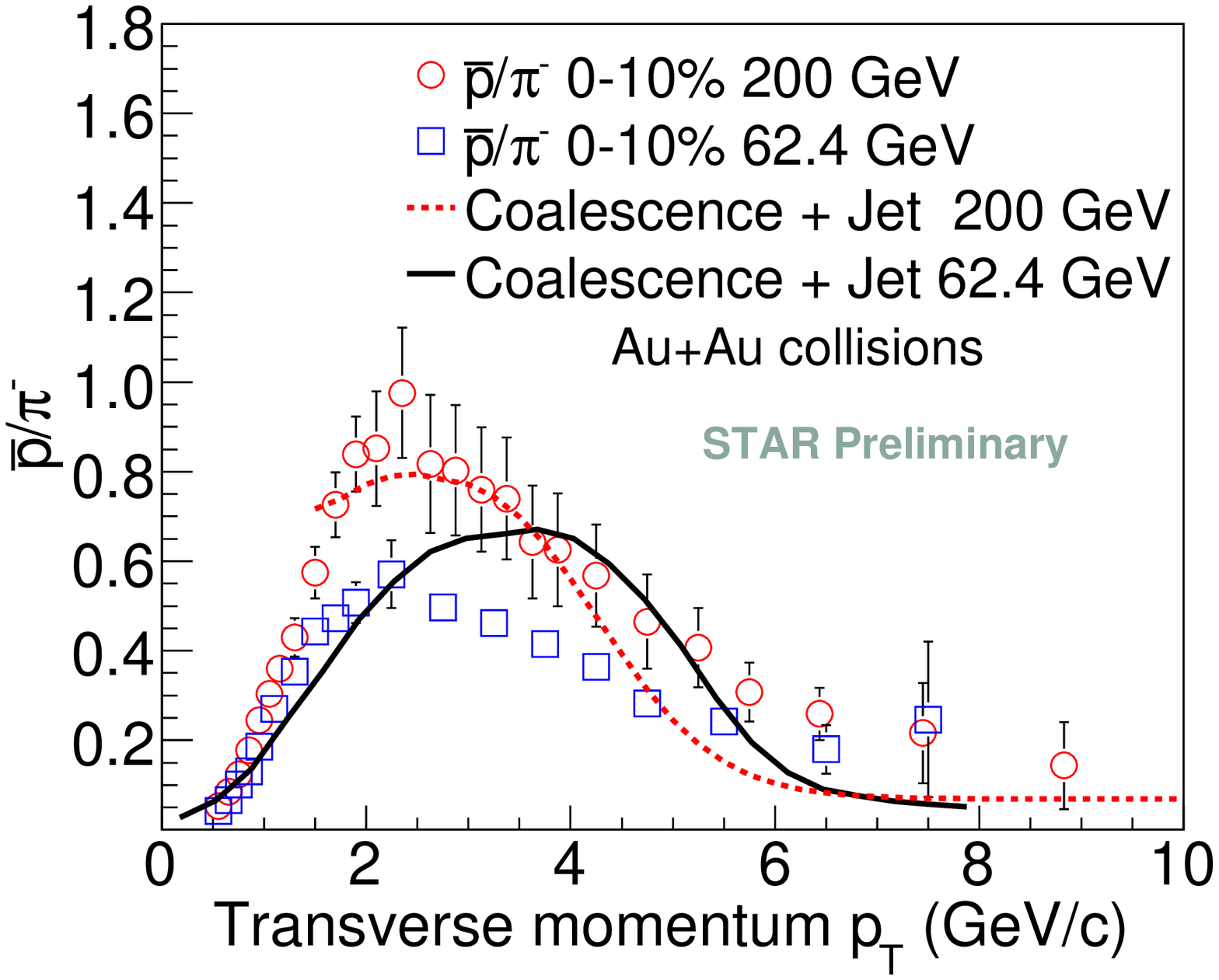}} %\vspace{-0.35cm}
\caption[]{(in color on line) The $\Lambda$/$K_{S}^{0}$,
$p/\pi^{+}$ and $\bar{p}/\pi^{-}$ ratios in central Au+Au
collisions at 62 and 200 GeV.} \label{ratio}
\end{figure}
The energy dependence of identified particle production provides a
tool to further test the coalescence and jet
quenching~\cite{jetquench}. Coalescence models predict a
higher(lower) $p/\pi^{+}$($\bar{p}/\pi^{-}$) ratio at 62 GeV
compared to 200 GeV at intermediate $p_T$~\cite{ko}.
Fig.~\ref{ratio} shows the $\Lambda/K_{S}^{0}$, $p/\pi^{+}$ and
$\bar{p}/\pi^{-}$ ratios as a function of $p_T$ in central Au+Au
collisions at 62 and 200 GeV~\cite{bmohanty}. At intermediate
$p_T$, we observe: $p/\pi^{+}$(62 GeV) $>$ $p/\pi^{+}$(200 GeV);
$\Lambda/K_{S}^{0}$ (62 GeV) $\sim$ $\Lambda/K_{S}^{0}$ (200 GeV)
and $\bar{p}/\pi^{-}$(62 GeV) $<$ $\bar{p}/\pi^{-}$(200 GeV).
Furthermore, we find the $\Lambda/K_{S}^{0}$, $p/\pi^{+}$ and
$\bar{p}/\pi^{-}$ ratios show similar peak positions and similar
shapes at these two energies. Also shown are the calculations from
coalescence + jet fragmentation models~\cite{fries,ko}. At
intermediate $p_T$, the calculations can reproduce the $p/\pi^{+}$
and $\bar{p}/\pi^{-}$ ratios at 200 GeV but can not reproduce the
ratios at 62 GeV. Compared to the data, the peak positions of the
ratios from model are shifted to higher $p_T$ with energy
decreasing. In general, at high $p_T$, jet fragmentation
models~\cite{fries,ko} can not reproduce the particle ratios.

\section{Study of Parton Energy Loss by Identified Baryon/Meson at High $p_T$}

At $p_T >$ 6 GeV/c, particle production is dominated by jet
fragmentation in Au+Au collisions. Identified particles at high
$p_T$ provide direct sensitivity to differences between quark and
gluon fragmentation. For example, the gluon contribution to
$\pi^{+}+\pi^{-}$ is around 50\% while that to $p+\bar{p}$ is
larger than 80\% at 6 $< p_T <$ 10 GeV/c in 200 GeV p+p collisions
~\cite{AKK,ppdAuPID}.
\begin{figure}[h] \centerline{
\includegraphics[width=0.5\textwidth]
{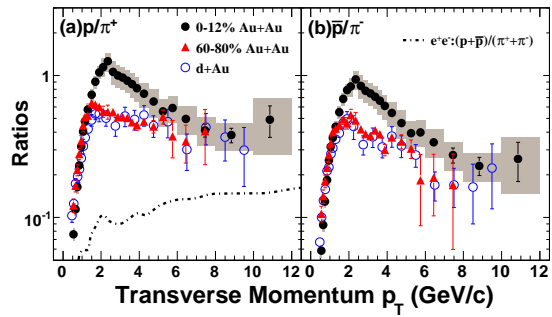} } \vspace{-0.35cm}
\caption[]{(in color on line) The $p/\pi^{+}$ and
$\bar{p}/\pi^{-}$ ratios in d+Au~\cite{ppdAuPID} and Au+Au
collisions~\cite{auauhighptPID}.} \label{ratiohighpt}
\end{figure}
The Casimir/color factors ($C_A$, $C_F$) of gluon and quark have
been determined in elementary $e^{+}+e^{-}$
collisions~\cite{colorlep}. It's found that $C_A/C_F$, the ratio
of the coupling strength of the triple-gluon vertex to that of
gluon bremsstrahlung from quarks is $\sim$ 2, which is in
agreement with the value expected from SU(3) QCD
theory~\cite{colorlep}. When traversing the hot and dense medium
created in the collisions, gluons lose more energy than
quarks~\cite{jetquench,starhighpt,xinnian:98}. Therefore, the
$\bar{p}/\pi$ ratio in central Au+Au collisions is expected to be
smaller than those in peripheral Au+Au, d+Au and p+p
collisions~\cite{xinnian:98}. Fig.~\ref{ratiohighpt} shows that
the $\bar{p}/\pi^{-}$($p/\pi^{+}$) ratios in central
Au+Au~\cite{auauhighptPID} and d+Au~\cite{ppdAuPID} collisions are
similar at $p_T >$ 5 GeV/c. This common degree of suppression of
protons and pions indicates that the partonic sources of pions and
protons have similar energy loss, which can not be explained by
simple color factor differences~\cite{xinnian:98}. One possible
solution is to consider additional processes in the interaction of
jets with the medium. For example, recently, the jet conversion in
the medium is considered and it is found that in order to
reproduce the data, a jet conversion rate much larger than
expected may be needed~\cite{jetconversion}.

\section{Summary}
In summary, we have presented a wealth of measurements from STAR
to study the properties of the partonic medium.

A common degree of suppression of protons and pions at high $p_T$
is observed. This new experimental phenomenon provides strong
constraints on the interaction of jets with the medium. The
interaction of jets with the medium leads to double peaked
structure on the away-side in di-hadron azimuthal correlations in
selected kinematic regions. The three particle correlations, in
the background-subtraction method, show evidence for conical
emission.

In the bulk, constituent interactions at the early stage transfer
the spatial anisotropy into significant momentum anisotropy in the
transverse plane. Significant fluctuations in $v_2$ have been
observed, which coincide with fluctuations in the initial
eccentricity. $v_2$ of identified particle $K_{S}^{0}$, $\Lambda$,
$\Xi$, $\Omega$, $K^*$ and $\phi$ at intermediate $p_T$ show
constituent quark number scaling, consistent with hadronization
through coalescence of constituent quarks from a collective
partonic system.

However, at intermediate $p_T$, there are several differences
observed between coalescence/recombination models and data. The
$p(\bar{p})/\pi$ ratios show similar peak positions at 62 and 200
GeV. A strong near-side correlation of $\Omega - h$ is observed
and the $\Omega/\phi$ ratio turns over at $p_T \sim$ 4 GeV/c.
These observations need further theoretical
exploration~\cite{hwaqm06}.

\end{document}